# An Algorithm for Unconstrained Quadratically Penalized Convex Optimization

## Steven P. Ellis

*Unit 42, NYSPI*
*1051 Riverside Dr.*
*New York, NY 10032*
*U.S.A.*
*e-mail:* `spe4@columbia.edu`

**Abstract:**
A descent algorithm, "Quasi-Quadratic Minimization with Memory" (QQMM), is proposed for unconstrained minimization of the sum, $F$, of a non-negative convex function, $V$, and a quadratic form. Such problems come up in regularized estimation in machine learning and statistics. In addition to values of $F$, QQMM requires the (sub)gradient of $V$.

Two features of QQMM help keep low the number of evaluations of the objective function it needs. First, QQMM provides good control over stopping the iterative search. This feature makes QQMM well adapted to statistical problems because in such problems the objective function is based on random data and therefore stopping early is sensible. Secondly, QQMM uses a complex method for determining trial minimizers of $F$.

After a description of the problem and algorithm a simulation study comparing QQMM to the popular BFGS optimization algorithm is described. The simulation study and other experiments suggest that QQMM is generally substantially faster than BFGS in the problem domain for which it was designed. A QQMM-BFGS hybrid is also generally substantially faster than BFGS but does better than QQMM when QQMM is very slow.

## 1. Introduction

### 1.1. Motivation

In order to make progress in solving a real world prediction problem of interest to the Department of Psychiatry at Columbia University, I have been led to develop a nonparameteric hazard function estimator for survival analysis. (This estimator will be described in another paper.) Fitting the nonparametric models required high dimensional numerical optimization for which I used the well regarded, but rather generic, BFGS optimizer. (See subsection 1.2 for

---

*This research is supported in part by United States PHS grant MH62185.

references.) I found that with BFGS the model fitting required hours. (Once it took 19 hours.)

Such long optimization times severely hampered development of the estimator so I cast about for an alternative to BFGS. My optimization problem is convex and convex optimization is an intensively studied problem (subsection 1.2). I was quite surprised, therefore, when I could not find an off-the-shelf, generic, unconstrained convex optimizer suitable to my problem. (More about this in subsection 1.2.) Consequently, I developed an apparently new, non-generic, optimizer, QQMM, that is the subject of this paper. I found that hazard estimation using QQMM generally took less than one hour, often substantially less. (See section 4 for an example.)

QQMM is designed for a specialized optimization problem class. However, the problem domain for which QQMM is appropriate is sufficiently broad that others might find QQMM useful.

### *1.2. Background*

A general strategy for nonparametric function estimation problems is as follows (28; 11; 29; 19).

1. Determine empirical risk functional. (Regard a minus log likelihood functional as a kind of risk functional.)
   - E.g., mean squared residual or misclassification rate.
2. To make optimization easier, if the risk functional is not already convex bound it above by a convex functional, $V$. Use $V$ in place of risk.
   - As $V$ is reduced, the original risk functional will tend to go down.
3. Only consider estimates in some "Reproducing Kernel Hilbert Space" (RKHS; (28)).
4. Regularize (15, p. 34) by adding a penalty that is a multiple of the squared norm in the RKHS.
5. Minimize the sum: $V + penalty$.

This strategy is advocated, at least for classification problems, in the following quote from (3, p. 337). "In our view, what is special about SVMs [Support Vector Machines] is the combination of the following ingredients: first and foremost, the use of positive definite kernels [item 3 in the preceding list]; then regularization via the norm in the associated reproducing kernel Hilbert space [item 4]; finally the use of a convex loss function [item 2] ..."

Thus, the idea is to minimize, in $h \in H_K$, where $H_K$ is an RKHS, a functional of the form.

$$F(h;Y) := F_Y(h) := V(h;Y) + \lambda \|h\|_K^2. \qquad (1)$$

Here, $Y$ is the data, $V$ is a convex functional that is larger than the empirical risk or minus log likelihood, $\lambda > 0$ is constant (the "complexity parameter"), and $\|\cdot\|_K$ is the norm of $H_K$. $F$ is the "objective function" (or "objective", for short). Call the expression $\lambda \|h\|_K^2$ the "penalty term." (See subsubsection 3.1.3 for an example.) There may also be side constraints on $h$, but in this paper we consider minimizing the functional (1) with no side constraints. We wish to find an approximation to a vector $h$ (a "minimizer") at which $F$ achieves its minimum value.

Thus, this paper concerns a class of convex optimization problems. Convex optimization is an important research area (e.g. (2; 4; 21; 16; 17)). However, it is difficult for most statisticians to tap into the products of this research area. To quote (14),

> ...there remains a significant impediment to the more widespread adoption of convex programming: the high level of expertise required to use it. With mature technologies such as [least squares], [linear programming], and [convex quadratic programming], problems can be specified and solved with relatively little effort, and with at most a very basic understanding of the computations involved. This is not the case with general convex programming. That a user must understand the basics of convex analysis is both reasonable and unavoidable; but in fact, a much deeper understanding is required. Furthermore, a user must find a way to transform his problem into one of the many limited standard forms; or, failing that, develop a custom solver. For potential users whose focus is the application, these requirements can form a formidable "expertise barrier" especially if it is not yet certain that the outcome will be any better than with other methods.

Grant *et al* (14) have developed, in MATLAB (The Mathworks, Natick, MA), a general purpose convex optimization modeling environment called "CVX" (http://www.stanford.edu/~boyd/cvx/). However, "CVX is not designed to handle problems such as [the survival analysis problem that prompted the development of QQMM]." –M.C. Grant, personal communication.)

As further evidence for the assertion of (14) note that apparently MATLAB's optimization toolbox does not come with functions for general convex optimization. Moreover there does not appear to exist a designated R (23) package for convex optimization.

QQMM is an easy to use "custom solver", in Grant *et al*'s terminology, for a large class of statistical problems. In addition to describing QQMM, I compare it to a popular "other method", namely BFGS (5; 12; 13; 24), (8, section 9.2), (22, Section 6.1). BFGS is a "quasi-Newton" method (22). Thus, BFGS requires only function values and gradients and with each iteration updates a matrix that it uses as one would use the inverse Hessian matrix in Newton's method. (See section 3.)

Moreover, an important assumption we make is that $V$ is bounded below by a known constant. Without loss of generality, we may take that constant to be 0. I.e., assume $V$ is non-negative. If $V$ is not bounded below by a known constant, replace $V$ by the point-wise maximum $\max\{V - c, 0\}$, for some constant, $c$. By taking $c$ sufficiently far below 0, this substitution may have little effect on the minimum of the objective.

In this paper I describe an apparently new algorithm "Quasi-Quadratic Minimization with Memory" (QQMM) for minimizing functions like $F$. QQMM was developed for a survival analysis application of the recipe presented in steps 1 – 5 above. (This application will be described in a future paper.) In the application to survival analysis the functional $V$ is time consuming to compute. QQMM may be best suited to functions like the survival analysis objective where $V$ is expensive to compute. (But see subsection 3.3.)

QQMM enjoys a good level of generality: It applies to a broad class of problems but does not suffer from the learning curve and inefficiency that come with greater generality.

Because of the kernel property (15, p. 146) minimizing (1), which *prima facie* is often an infinite dimensional problem, often reduces to a, perhaps large, finite dimensional unconstrained problem. Call the dimension of the reduced problem $d$. (See subsubsection 3.1.3 for an example.)

Kernel-based methods (1) provide a motivation for the optimization problem we are considering. But a simpler way to describe the problem is as follows. Start with a known symmetric, positive definite quadratic form $Q$ (25) on a $d$-dimensional real linear space, $H$. Thus, for a given basis on $H$, there is a $d \times d$ symmetric, positive definite matrix, $\mathbf{Q}$, such that if $h \in H$ has coordinates $\mathbf{x} := (x_1, \ldots, x_d)$ relative to the basis then $Q(h) = \mathbf{x}\mathbf{Q}\mathbf{x}^T$, where "$T$" indicates matrix transposition. Consider minimizing, in $h$, the function

$$F(h) := V(h) + Q(h), \quad h \in H, \tag{2}$$

for some non-negative convex function, $V$, on $H$.

QQMM is an algorithm for minimizing (2). QQMM requires the inverse of a matrix, $\mathbf{Q}$, defining $Q$ so $\mathbf{Q}$ has to be strictly positive definite. (See subsection 2.2 for an explanation and discussion.)

*Remarks:*

1. The quadratic part does not have to come from the penalty term. E.g., one should be able to use QQMM for lasso regression (26), (15, p. 64) in which $V$ provides the quadratic part.
2. The requirement that $V$ be convex can be relaxed. If $V$ is the difference between two convex functions then the minimization of $F$ can be reduced to iterative application of QQMM to a sequence of convex $V$'s (7, p. 322), (20).

## *1.3. QQMM*

The left panel of figure 1 shows QQMM in action on a convex function of two variables. Numbers 1 through 9 label the positions at which and the order in which the function was evaluated during the iterative search for the minimum. (Positions 7, 8, and 9 are crowded together in the middle.) Twice, in moving from position 4 to position 5 and in moving from position 7 to position 8, QQMM overshot and actually increased the objective function. In those cases the algorithm backtracked. (The right hand panel of figure 1 will be explained in subsection 2.4.)

QQMM requires the functions $V$ and $Q$ and a function, $g$, that returns subgradients of $V$ (subsection 2.1; subsubsection 2.2.1).

$F_Y$ in (1) depends on $Y$, a data sample. If $Y_1$ and $Y_2$ are two different samples, the location of the minimum of $F_{Y_1}$ will probably not be the location of the minimum of $F_{Y_2}$. For that reason there is no need to approximate the minimizer of $F$ very precisely. For model selection or even more for resampling (9), computational speed is an issue. An advantage of QQMM is that as it searches for a minimum it computes ever more accurate lower bounds on $F$. One can use this lower bound in a stopping rule for the search. For example, one could specify that the search stop when the objective and lower bound differ by less than 1% (subsection 2.4).

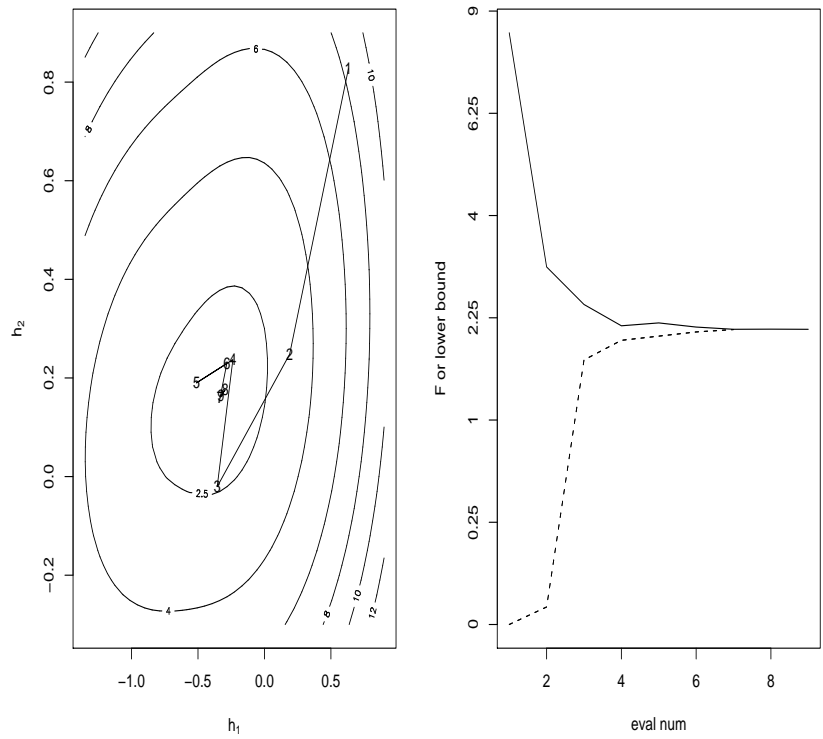

Fig 1. *QQMM search path with a random convex function in two variables. Left panel shows contour plot of objective function and the 9 trial minimizers numbered in order of evaluation and connected by lines. Right panel shows function values at the trial minimizers (solid curve) and lower bounds (dashed). Horizontal axis is evaluation number. (Vertical axis is on square root scale.)*

The objective function for the survival analysis algorithm for which QQMM was developed is expensive to compute. Therefore, an especially important goal in developing QQMM was that it converge without many function evaluations. This is accomplished through two means. First is the aforementioned ability to stop early. Another way that QQMM is designed to require few evaluations is that it performs a fair amount of computation to carefully determine trial minimizers (subsections 2.2 and 2.3). This is cost effective because, as mentioned above, QQMM was devised to minimize $F$ for a specific form of $V$ that is expensive to evaluate. This extra computational effort may be counter productive if computing $V$ is cheap. (But see subsection 3.3.)

The cumulative amount of computational overhead for QQMM is approximately proportional to the square of the number of function evaluations and memory requirements are roughly linear in the number of evaluations. If QQMM needs many function evaluations and $d$ is large this burden can be important.

The distribution of the number of objective function evaluations needed by QQMM seems to have a long right hand tail. E.g., if the quadratic form $Q$ has small eigenvalues, say, if $\lambda$ in (1) is small, then QQMM can be quite slow. A way to correct that is "BFGS rescue" (subsection 2.5) in which the BFGS algorithm takes over if QQMM runs too long. However, "most of the time" QQMM seems not to need rescue (section 3).

QQMM involves matrix operations so that in practice there is a limit to how large $d$ can be. I have successfully used QQMM with $d$ as large as 5,000.

I have implemented QQMM in R. (The source code is available from me.) The algorithm is presented in section 2. In section 3 QQMM is compared to BFGS in kernel-based, nonparametric $L^{3/2}$-regression with thousands of simulated datasets. A summary and further discussion are presented in section 4. (Part of this paper appeared as (10).)

Since QQMM is rather complex, I make no attempt here to analyze it theoretically. Moreover, the notion of "stopping early" appears to be a rather nontraditional notion in the theoretical analysis of optimization algorithms.

## 2. The Algorithm

QQMM proceeds interatively beginning with a starting vector, $h := startVec$. The algorithm generates a sequence of "trial minimizers" $h_{(1)}, h_{(2)}, \ldots \in H$ until some criteria are met (subsection 2.4). Then it halts. QQMM is a descent method (18, p. 427) so that after each iteration the value of $F$ is strictly reduced. In fact, after each iteration $F$ is "sufficiently" decreased (subsection 2.3). Not all trial minimizers will decrease the value of $F$ sufficiently. The trial minimizers that do decrease the value of $F$ sufficiently are "iterates". (The starting value of $h$ is, trivially, considered to be an iterate.) Thus, an iteration is complete when an iterate is found, but $F$ may have to be evaluated at several trial minimizers in order to find an iterate. When QQMM halts it returns (among other things) the last iterate (unless it halts because the evaluation count limit has been reached in which case it returns the best $h$ it has found, iterate or not).

### 2.1. Top level description

Here, I describe, somewhat abstractly, the general structure of QQMM. Let $\langle \cdot, \cdot \rangle$ be an inner product on $H$. If $h_0 \in H$, let $g(h_0) \in H$ be a "subgradient" of $V$ at $h_0$ (21, p. 126). This means the following. Let $p_{h_0}$ be the affine function

$$p_{h_0}(h) := V(h_0) + \langle h - h_0, g(h_0) \rangle \qquad (h \in H). \tag{3}$$

Then $p_{h_0} \leq V$ pointwise. This is only possible in general because $V$ is convex. Typically, the actual gradient $\nabla V(h_0)$ at $h_0$ will exist (16, Theorem 4.2.3, p. 189). If so, $g(h_0) = \nabla V(h_0)$.

Let $h \in H$. There is an operation that takes the triple $\big(h, F(h), g(h)\big)$ to a global underestimator (4, p. 69) of $F$, i.e., a function that is nowhere larger than $F$ (subsubsection 2.2.1). Call that global underestimator $minorant(h)$. As the algorithm proceeds it in effect accumulates the minorants in a list. Call that list "$minorants$". It is because QQMM accumulates information as it proceeds that the second "M", for "Memory", is included in the name of the algorithm.

Let "$\mathbb{R}$" = "reals". Recall that $V$ is non-negative. Let $startVec \in H$. $startVec$ will be a vector of starting values for the algorithm. In the following $h_1$ will denote the current iterate.

$minorants \leftarrow empty\ set$; $LargstLwrBndSoFar \leftarrow 0 \in \mathbb{R}$;
$eval.count \leftarrow 0$; $h_1 \leftarrow startVec$;
   While $eval.count \leq max.evals$;
      If criteria for **stopping** are met then break out of loop;
      If $F(h_1)$ and $g(h_1)$ have not already been evaluated,
         then evaluate $F(h_1)$ and $g(h_1)$ and
            **compute minorant**$(h_1)$;
         $eval.count \leftarrow eval.count + 1$;
         Append $minorant(h_1)$ to $minorants$;
      Use $minorants$ and $LargstLwrBndSoFar$ to
         **find next trial minimizer**, $h_2$,
         and update $LargstLwrBndSoFar$;
      Evaluate $F$ and $g$ at $h_2$ and append $minorant(h_2)$ to $minorants$;
      $eval.count \leftarrow eval.count + 1$;
      While $F(h_2)$ is not **sufficiently decreased** from $F(h_1)$
               and $eval.count \leq max.evals + 1$;
         Compute new $h_2$ on the line $\{t(h_2 - h_1) + h_1 : t \in \mathbb{R}\}$;
         Compute $F(h_2)$ and $g(h_2)$;
         $eval.count \leftarrow eval.count + 1$;
      End while;
      Append $minorant(h_2)$ to $minorants$;
      (*) If $F(h_2) < F(h_1)$ then $h_1 \leftarrow h_2$;
   End while;
Return $h_1$;

*Remarks:*

1. *minorants* does not have to be initialized to the empty set. It could be initialized to a list generated by a previous run of the algorithm, for example.
2. If $F(h_2)$ is sufficiently decreased from $F(h_1)$ then, of course, $F(h_2) < F(h_1)$. It is necessary to check $F(h_2) < F(h_1)$ at the step labeled (*) because the algorithm might break out of the sufficient decrease loop due to the evaluation limit having been reached.

To complete the description of the algorithm I must spell out, first, what the minorants are and how they are used to generate the next trial search vector (and how $LargstLwrBndSoFar$ is updated), second, what is meant by "sufficient decrease" and how a new $h_2$ is chosen on the line through $h_1$ and $h_2$, and, third, when to stop. These are described in subsections 2.2, 2.3, and 2.4, respectively. In subsection 2.5 I describe a hybrid of QQMM and BFGS.

### 2.2. Minorants

In this subsection I give details on step "**find next trial minimizer**" in the algorithm outline in subsection 2.1. The naive idea is to use at each iteration all the information about $F$ accumulated so far to bound $F$ below by another convex function, $\mathcal{F}$. Then one chooses the minimizer of $\mathcal{F}$ to be the next trial minimizer of $F$. However, that idea is impractical. Instead, QQMM minimizes multiple convex functions bounding $F$ below and chooses the one whose minimum value is largest.

Suppose we have accumulated a list of trial minimizers. Let $h_1$ be the current iterate. Let $LargstLwrBndSoFar$ be the current value of the lower bound to $F$. If the stopping criteria (subsection 2.4) are not satisfied, QQMM needs to search for $h_2 \in H$ at which $F$ is "sufficiently decreased" compared to $F(h_1)$ (subsection 2.3). (Once such an $h_2$ is found, QQMM sets $h_1 \leftarrow h_2$ and, providing the evaluation limit, $max.evals$, has not been reached, repeats the process.) Iteration should, of course, stop if $F(h_1) = LargstLwrBndSoFar$. So assume $F(h_1) > LargstLwrBndSoFar$.

### 2.2.1. Computing minorants

Recall that $V$ is non-negative, let $h_0 \in H$, and recall the definition, (3), of $p_{h_0}$. Then, again because $V$ is convex, the function

$$q_{h_0}(h) := q\big(h; h_0, V(h_0), g(h_0)\big) := \max\big\{p_{h_0}(h), 0\big\} + Q(h), \quad h \in H,$$

is a convex global underestimator of $F$. (The function $q_{h_0}$ is what is called $minorant(h_0)$ in subsection 2.1.) Call $q_{h_0}$ a "quasi-quadratic" function. (Figure 2 shows examples. The "QQ" in "QQMM" stands for "quasi-quadratic".) It is easy to minimize $q_{h_0}$, and the minimum value is a lower bound on $F$ that is at least 0.

### 2.2.2. Using minorants to compute a trial minimizer

Suppose the current list of minorants is $minorants = \{q_{h_{(1)}}, \ldots, q_{h_{(n-1)}}\}$. Let $h_{(n)} := h_1$ be the last iterate. QQMM appends $q_{h_{(n)}}$ to $minorants$. (More precisely, QQMM maintains lists of trial minimizers and corresponding values and subgradients of $V$. "Appending $h_{(n)}$ to $minorants$" means appending $h_{(n)}$, $V(h_{(n)})$, and the subgradient $g(h_{(n)})$ to the appropriate lists. I.e., QQMM stores the "raw bundle", (17, p. 228).) The (pointwise) least upper bound

$$\mathcal{F}_n := \max\{q_{h_{(1)}}, q_{h_{(2)}}, \ldots, q_{h_{(n-1)}}, q_{h_{(n)}}\}$$

is a convex global underestimator of $F$. ($\mathcal{F}_n$ is a "model" of the convex function $F$, (21, p. 157), (22, Section 9.2).) In theory we could find the next trial minimizer by minimizing $\mathcal{F}_n$.

However, if $n$ is large then minimizing $\mathcal{F}_n$ could be harder than minimizing $F$. Instead, QQMM proceeds as follows. For $i = 1, 2, \ldots, n-1$ let $q_{h_{(i)}, h_1}$ be the pointwise maximum

$$q_{h_{(i)}, h_1}(h) := \max\big\{q_{h_{(i)}}(h), q_{h_1}(h)\big\} = \max\big\{q_{h_{(i)}}(h), q_{h_{(n)}}(h)\big\}, \quad h \in H.$$

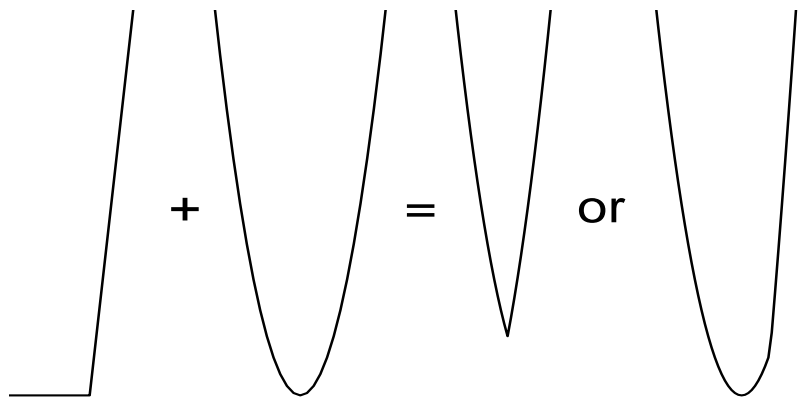

FIG 2. *Two quasi-quadratic functions on the line. A quasi-quadratic function is a "hinge" plus a quadratic. In general, hinges do not fold at the origin. (In higher dimensions hinges fold along planes that in general do not pass through the origin.) Where the hinge folds influences the shape of a quasi-quadratic function, as shown here.*

This function is a convex global underestimator to $F$ that is easy to minimize in closed form. (In the current implementation minimizing $q_{h,h_1}$ requires the inverse of a matrix, $\mathbf{Q}$, defining $Q$. If $\mathbf{Q}$ is ill-conditioned one can improve the conditioning by replacing $\mathbf{Q}$ by the matrix obtained by building up the diagonal of the matrix. This is illustrated in subsubsection 3.1.3.)

Let $i = j$ be the index in $\{1, 2, \ldots, n-1\}$ such that the *minimum* value of the function $q_{h_{(i)},h_1}(h)$ is the *largest*, i.e., no smaller than the minimum value of any $q_{h_{(i)},h_1}$ ($i = 1, \ldots, n-1$). (So $h_{(j)}$ has a "maximin" property.) Let $h^{mm}$ be the vector at which $q_{h_{(j)},h_1}$ achieves its minimum. Denote the minimum value, $q_{h_{(j)},h_1}(h^{mm})$, by $m$. Let $LargstLwrBndSoFar$ be the largest (constant) lower bound to $F$ that QQMM has found so far. (Do not confuse $LargstLwrBndSoFar$, a lower bound on $F$ that changes during the course of the calculation, with 0, a fixed lower bound on $V$.) If $m \geq LargstLwrBndSoFar$ then the next trial minimizer will be based upon $\tilde{h}_2 := h^{mm}$ and QQMM sets

$$LargstLwrBndSoFar \leftarrow m.$$

However, if $m < LargstLwrBndSoFar$ then QQMM performs a "shoreline" operation to arrive at $\tilde{h}_2$. (Think of $LargstLwrBndSoFar$ as "sea level". See subsubsection 2.2.3.)

In general, QQMM does not use $\tilde{h}_2$ as a trial minimizer $h_2$. It sets

$$h_2 \leftarrow h_1 + daring \times (\tilde{h}_2 - h_1). \tag{4}$$

In the numerical experiments described in this paper *daring* is just the constant 0.7 (subsubsection 3.2.1), however, one might let *daring* be, say, a slowly decreasing function of the iteration number.

Before replacing $h_1$ by $h_2$ defined by (4) QQMM tests to see whether $F(h_2)$ represents a "sufficient decrease" from $F(h_1)$. This process is described in subsection 2.3. If $F(h_2)$ does *not* represent a "sufficient decrease" from $F(h_1)$, then $h_2$ is replaced by $h_1 + \delta(h_2 - h_1)$ for some user-specified constant $\delta \in (0, 1)$. For the calculations made for this paper, $\delta := 1/6$ (subsubsection 3.2.1).

### 2.2.3. "Shoreline operation"

This means the following. Suppose $m < LargstLwrBndSoFar$. Let $h_{1*}$ be the vector at which $q_{h_1}$ is minimized. Then by definitions

$$q_{h_1}(h_{1*}) \le q_{h_1}(h^{mm}) \le q_{h_{(j)}, h_1}(h^{mm}) = m$$
$$< LargstLwrBndSoFar < F(h_1) = q_{h_1}(h_1).$$

Hence, since $q_{h_1}$ is convex there exists a unique $t \in (0, 1)$ at which

$$q_{h_1}\bigl[t(h_{1*} - h_1) + h_1\bigr] = LargstLwrBndSoFar.$$

QQMM sets $\tilde{h}_2 := t(h_{1*} - h_1) + h_1$.

### 2.3. Sufficient decrease

If a trial minimizer, $h_2$, were accepted as an iterate if it merely reduced the size of the objective compared to the current best value, $F(h_1)$, the objective could conceivably decrease unacceptably slowly from one iteration to the next. To help prevent this, QQMM requires a "sufficient decrease" (22, p. 33) in $F$.

Figure 3 shows how the sufficient decrease test works. The solid curves are graphs of $F$ along the line passing through $h_1$ and $h_2$. The dotted horizontal lines are at a height equal to $F(h_1)$. The values $h_{2a}$ through $h_{2f}$ are possible positions of $h_2$ along the line. The sloping dashed lines pass through $\bigl(h_1, F(h_1)\bigr)$ and have slopes proportional to the slopes of $F$ at $h_1$. (The constant of proportionality – call it $\sigma$ – is a tuning constant of the algorithm; see subsubsection 3.2.1. For the calculations presented in this paper I used $1/6$, but in the figure $\sigma \ne 1/6$.) If $F(h_2) \ge F(h_1)$ as at $h_2 = h_{2d}$, then there is no decrease, so certainly not a sufficient decrease. We have $F(h_{2c}) < F(h_1)$ but the decrease in $F$ at $h_2 = h_{2c}$ is not deemed sufficient because (1) the point $\bigl(h_{2c}, F(h_{2c})\bigr)$ lies above the dashed line and (2) at $h_2 = h_{2c}$ the function $F$ is increasing with increasing distance from $h_1$. The point $\bigl(h_{2f}, F(h_{2f})\bigr)$ also lies below the dotted and above the dashed line but there the function $F$ is *de*creasing with increasing distance from $h_1$. Under those conditions the decrease in $F$ at $h_2 = h_{2f}$ is defined to be sufficient. At $h_2 = h_{2a}$, $h_{2b}$, and $h_{2e}$, the point $\bigl(h_2, F(h_2)\bigr)$ lies below the dashed line and so the decrease is defined to be sufficient at those points. So even though $F(h_{2c}) < F(h_{2a})$, $F$ exhibits sufficient decrease at $h_2 = h_{2a}$ but not at $h_2 = h_{2c}$.

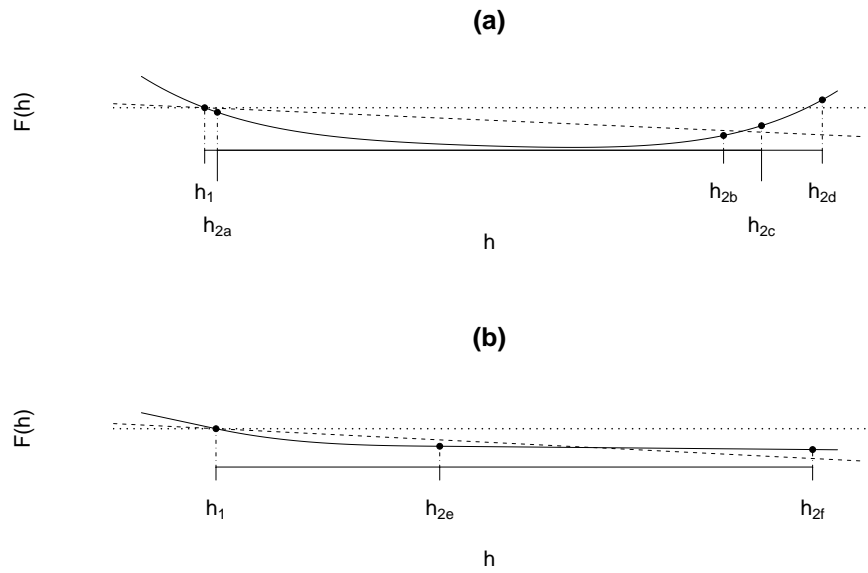

FIG 3. *Sufficient decrease. Curves are graphs of $F$ along the line passing through $h_1$ and $h_2$. $h_{2a}$ through $h_{2f}$ are possible positions of $h_2$ along the line. Sloping dashed lines pass through $\bigl(h_1, F(h_1)\bigr)$. Horizontal dotted lines also pass through $\bigl(h_1, F(h_1)\bigr)$. $F$ shows sufficient decrease at $h_{2a}$, $h_{2b}$, $h_{2e}$, and $h_{2f}$, but not at $h_{2c}$ or $h_{2d}$.*

If $F$ does not exhibit sufficient decrease at $h_2$ then $h_2$ is replaced by

$$h_2 \leftarrow h_1 + \delta(h_2 - h_1).$$

Providing $eval.count \leq max.evals + 1$, a new test for sufficient decrease is then performed. In the computations done for this paper, I took $\delta := 1/6$ (subsubsection 3.2.1).

## *2.4. Stopping rules*

An important feature of QQMM is that it allows convenient stopping rules. That is because in the course of its operation, it produces a nondecreasing sequence of constant lower bounds, $LargstLwrBndSoFar$, to $F$. So as the computation proceeds one can, for example, bound the relative error by known quantities:

$$\frac{F(h_1) - LargstLwrBndSoFar}{LargstLwrBndSoFar} \geq \frac{F(h_1) - \min F}{\min F}. \tag{5}$$

Therefore, a natural stopping rule is to stop when

$$\frac{F(h_1) - LargstLwrBndSoFar}{LargstLwrBndSoFar} < \epsilon, \tag{6}$$

where $\epsilon > 0$ is a prespecified constant. Then one knows that the relative error in the output is no greater than $\epsilon$. (As a precaution, one should also stop when $g(h_1)$, where $g(h_1)$ is a subgradient of $F$ at $h_1$, is very close to 0, when $F(h_1)$ is very close to 0, and when $h_1$ is very close to one of the vectors $h_{(1)}, \ldots, h_{(n-1)}$.) The right panel of figure 1 shows an example. In this paper I use $\epsilon := 0.01$. The horizontal axis is evaluation number (of $F$). The solid curve is the value of $F$ produced by the corresponding evaluation. The dashed line shows, at evaluation $i$, the corresponding value of $LargstLwrBndSoFar$.

Note that, obviously, when stopping rule (6) halts QQMM the real relative error will be *strictly* less than $\epsilon$. But on top of that, in practice the inequality (5) will be strict. This leads one to suspect that in practice the true relative error at the final location of $h_1$ will tend to be considerably less than $\epsilon$. Panel (c) of figure 5 shows true relative errors from the simulation study presented in section 3. In that study $\epsilon := 0.01$. The actual relative errors tended to be far smaller than that.

As indicated in subsection 2.1 another way the algorithm halts is if the number of function evaluations exceeds the limit $max.evals$. (There might be another evaluation or two before QQMM comes to a full stop.)

Having good control over when the iteration stops is particularly important in statistical applications where the objective function $F$ depends on random data. Since $F$ is random it does not make sense to strive for high precision in approximating its minimizer.

### *2.5. "BFGS rescue"*

QQMM usually closes in on the minimum of the objective function rapidly but is sometimes very slow to converge. I.e., the distribution of the number of evaluations until convergence has a long right hand tail. To address this problem, one can employ a "BFGS rescue". This means the following procedure. First, run QQMM for a limited number of evaluations. *If* QQMM does not converge, then perform a limited number of iterations of BFGS from where QQMM left off. Denote this procedure by "QQMM$_{rescue}$". See subsection 3.2 for a specific example.

## 3. Numerical Example: Nonparametric $L^{3/2}$-regression

### *3.1. Simulation*

#### *3.1.1. Strategy*

I carried out a simulation study of QQMM and BFGS. This means comparing QQMM and BFGS on many different problems. However, QQMM is designed for "expensive" $V$'s, i.e., $V$'s that are time consuming to compute. Many runs of QQMM an BFGS on problems with expensive $V$'s would take prohibitively long. (But I report anecdotally on a BFGS/QQMM comparison on an expensive $V$ problem in section 4.) As a way out of this quandary just note that, obviously, the reason a QQMM or BFGS application to a problem with an expensive $V$ might take a long time is that it takes time to *evaluate V*. I.e., number of function evaluations, always a proxy for optimization time, is an especially good proxy when $V$ is expensive to compute. This means we might get insight into the running times of QQMM and BFGS on problems with expensive $V$ by doing a large simulation study for QQMM and BFGS on problems with *in*expensive $V$'s and taking, instead of running time, *number of evaluations* as the outcome measure of interest. This is the strategy I adopted.

The mean number of evaluations is related to total amount of time needed for optimizations to finish. So the *mean* number of evaluations required by optimizer (and not, e.g., the median) is the crucial parameter. (It turns out that for comparison with BFGS, the median is, if anything, a more favorable parameter for QQMM than is the mean. But the mean is the one of more practical relevance.)

The current implementation of QQMM and `optim`'s implementation of BFGS count "evaluations" differently. For QQMM an evaluation means computing the value of $F$ ("value evaluation") *together with* the subgradient, $g$, ("gradient evaluation"). Each time this happens *eval.count* is incremented by 1. BFGS counts value evaluations and gradient evaluations separately. By "total number of evaluations" by BFGS I will mean the number of value evaluations *plus* the number of gradient evaluations. The "total number of evaluations" by QQMM will mean a comparable number (essentially twice the final value of *eval.count*). (In subsection 3.3 I briefly discuss system time comparisons.)

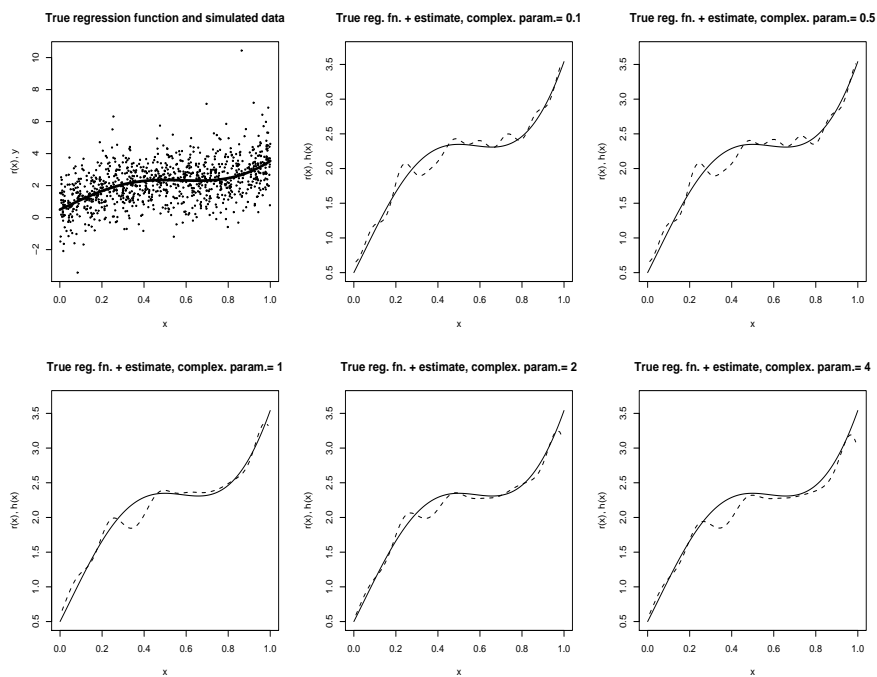

Fig 4. *Upper left panel: The function (7) superimposed on 1000 data points generated from a model with regression function (7). The other five panels show the same function (solid lines) together with kernel estimates (dashed lines) with a variety of complexity parameter values and based on the data set shown in the upper left panel. (Upper left plot uses a different vertical scale than do the other plots.)*

### 3.1.2. Regression problem

We compare QQMM$_{rescue}$ with BFGS on a simulated nonparametric regression problems. In each case, the "true" regression function is

$$r(x) := 0.5 + x + 3x^2 + \sin(5x), \quad 0 < x < 1 \qquad (7)$$

(27). The graph of this function is shown in figure 4.

In each run, the simulated data consisted of 1000 data points. (While there is just one predictor variable, as explained in subsubsection 3.1.3, there are $d := 1000$ variables for optimization.) The predictors, $x_1, \ldots, x_{1000}$ are uniformly distributed on (0,1). The errors are independent of the $x_i$'s and independent and identically distributed (i.i.d.) with $t$ distribution with 7 degrees of freedom. Let

$$y_i := r(x_i) + error_i, \quad i = 1, \ldots, 1000.$$

(See figure 4 for a plot of one such data set.)

*3.1.3. Estimation*

Employ kernel-based $L^{3/2}$-regression with Gaussian kernel

$$K(x, x') := \exp\left\{-100|x - x'|^2\right\}$$

(6). So in this case

$$F(h) = F(h; \mathbf{x}, \mathbf{y}) = \sum_{i=1}^{1000} |y_i - h(x_i)|^{3/2} + \lambda \|h\|_K^2. \tag{8}$$

Here, $\|h\|_K$ is the norm in the RKHS, $H_K$, corresponding to $K$ and $\lambda$ is the "complexity parameter." By the "kernel property" (15, p. 146), the infinite dimensional problem of minimizing $F$ in (8) over $H_K$ reduces to the finite dimensional problem specified by taking

$$h(x) := \sum_{i}^{1000} h_i\, K(x, x_i),$$

where $h_1, \ldots, h_{1000} \in \mathbb{R}$ and, using the symbol $\mathbf{h}$ to denote the vector $(h_1, \ldots, h_{1000})$, we have

$$\|h\|_K^2 = \mathbf{h}\,\mathbf{G}\,\mathbf{h}^T.$$

Here, $\mathbf{G}$ is the $1000 \times 1000$ "Gram matrix" (6, p. 169), whose $(i,j)^{th}$ entry is $K(x_i, x_j)$. Thus,

$$Q(h) = \lambda \|h\|_K^2 = \lambda\,\mathbf{h}\,\mathbf{G}\,\mathbf{h}^T. \tag{9}$$

(I chose $L^{3/2}$ regression instead of $L^2$ regression because kernel-based $L^2$ regression has a closed form solution (6, section 6.2.2) so does not require numerical optimization. Another standard choice would be kernel-based $L^1$ (absolute deviation) regression, but the $L^1$ criterion $(V)$ is not everywhere differentiable and BFGS is designed for differentiable functions.)

It turns out that for the simulated data sets described in subsubsection 3.1.2 the Gram matrix, $\mathbf{G}$, is often poorly conditioned. But to compute the minima of the $q_{h_{(i)}, h_1}$'s (subsubsection 2.2.2) QQMM requires the inverse of the matrix, $\lambda \mathbf{G}$, of $Q$. As a work around, for each randomly generated data set, in (9) the matrix $\mathbf{G}$ is replaced by $\mathbf{G} + \delta \mathbf{I}$, where $\mathbf{I}$ is the $1000 \times 1000$ identity matrix and $\delta$ equals 0.005 times the mean of the diagonal entries of $\mathbf{G}$. (Of course, the mean of the diagonal entries equals the mean of the eigenvalues. In fact, the diagonal of $\mathbf{G}$ consists of 1's, so each diagonal entry gets replaced by 1.005.)

Figure 4 shows estimates of $r$ for various values of $\lambda$ and one simulated data set. The estimates with $\lambda := 0.1$ and $0.5$ show "overfitting". The estimates with $\lambda := 1$ or $2$ tracks the target, $r$, rather well. The estimate with $\lambda := 4$ exhibits "underfitting".

(In practice one might use cross-validation to assess the quality of estimates. This paper is concerned with the computational, not statistical, aspects of this estimation problem. In the course of model tuning one will normally fit models, and so perform optimization, with $\lambda$ too large and too small as well as near optimum $\lambda$.)

### 3.2. Compare QQMM and BFGS in computing $L^{3/2}$-estimates

For each value of $\lambda := 0.1, 0.5, 1, 2,$ and $4$, I generated 1000 independent data sets as above. For each such data set I also generated an independent random vector, **z**, of 1000 i.i.d. standard normal variates. (**z** was generated independently for each of the 1000 data sets.) Then I minimized $F$ using BFGS and QQMM$_{rescue}$ for each of the 1000 data sets, in each case starting from the corresponding **z**. As noted above, we are interested in the mean number of evaluations used by each optimization method. Of course, an acceptable level of accuracy also required.

I used BFGS with an iteration limit of 1,000,000 ("BFGS$_{1,000,000}$") as the standard. (Do not be put off by the "1,000,000". Actually, BFGS$_{1,000,000}$ never needed more than 250 evaluations.) I compared BFGS$_{1,000,000}$ to QQMM$_{rescue}$ with

- Stopping threshold $\epsilon := 0.01$.
- At most $\approx max.evals := 110$ QQMM evaluations.
- Followed by an at most $max.iter := 35$ iteration BFGS rescue, *if needed*.

#### 3.2.1. Choosing the tuning constants

First, I describe how the parameters *max.evals* and *max.iter* were arrived at. By varying two parameters one can hope to achieve two goals. The first goal was comparability between QQMM$_{rescue}$ and BFGS. The optimizers BFGS and QQMM can be tuned in many different ways. In order to make the comparison meaningful we need to use comparable versions. In the simulations *max.evals* and *max.iter* were chosen so that for $\lambda := 0.1, 0.5,$ and $1$, the maximum number of evaluations QQMM$_{rescue}$ needed was never smaller than the maximum number needed by BFGS$_{1,000,000}$. The idea is to rule out the possibility that a lower mean number of evaluations for QQMM$_{rescue}$ compared to BFGS$_{1,000,000}$ would simply be due to QQMM$_{rescue}$ being allowed fewer evaluations than the number needed by BFGS$_{1,000,000}$. (Of course, the maximum number of evaluations is a rather unstable statistic. This makes the tuning rather approximate.) That parity broke down for $\lambda \geq 2$, where QQMM needed far fewer evaluations. For $\lambda := 2$, BFGS with iteration limit of only 10 ("BFGS$_{10}$") was also run on the same simulated problems (subsubsection 3.3.1). The second goal driving the choice of *max.evals* and *max.iter* was that the relative errors for QQMM$_{rescue}$ be of tolerable size in the simulations.

The choice of stopping threshold $\epsilon := 0.01$ was a judgment that a relative error less than 1% would be acceptable for most statistical purposes. (As we will see in subsection 3.3 the actual relative errors were usually far smaller than 0.01.) The other internal tuning constants in QQMM, *viz. daring* (subsubsection 2.2.2) and the tuning constants that define sufficient decrease, *viz.*, $\sigma$ and $\delta$ (subsection 2.3) were chosen as follows. A simulation study was carried out in which tens of thousands of random two-dimensional optimization problems (not shown) such as that shown in figure 1 were generated and optimized using QQMM (without rescue) with various choices of the tuning constants. A large regression model of evaluation count vs. tuning constants was fitted to the results. The tuning constants *daring*, $\sigma$, and $\delta$ were chosen to approximately minimize the fitted regression function. (Thus, the tuning constants were chosen based on problems different from the regression problems used for testing QQMM.)

## 3.3. Results

The results are summed up in figure 5. We see from panel (b) that for $\lambda := 0.1$, QQMM$_{rescue}$ required, on average, about 13% more evaluations than did BFGS$_{1,000,000}$. Again with $\lambda := 0.1$, QQMM$_{rescue}$ required rescue every time (panel (d)). As $\lambda$ increased the average number of evaluations required by QQMM decreased in absolute terms (panel (a)) and relative to the average number required by BFGS$_{1,000,000}$ (panel (b)) and and QQMM required "rescue" less often (panel (d)). By the time $\lambda$ reached 2, QQMM$_{rescue}$ required only a little less than one third as many evaluations as did BFGS$_{1,000,000}$ and did not require "rescue" at all.

Figure 5 (b) also shows what happened when no rescue was made. Then with $\lambda := 0.1$, QQMM required 60% more evaluations than did BFGS$_{1,000,000}$. However, for $\lambda := 0.5$ or 1, QQMM was actually a little faster without rescue than with. The additional evaluations needed by QQMM$_{rescue}$ for these $\lambda$ values can be thought of as "insurance premiums" to pay for a reduction in needed evaluations at $\lambda := 0.1$.

Define acceptable accuracy to mean relative error (compared to BFGS$_{1,000,000}$) less than 1%. (Recall that the accuracy target, $\epsilon$ – see subsection 2.4 – had been set to 1%.) In that case, the accuracy of QQMM$_{rescue}$ was almost always acceptable. When $\lambda := 0.1$ the maximum relative error in the 1,000 problems was 1.4%. The $95^{th}$ percentile of relative error when $\lambda := 0.1$ was 1%. For $\lambda := 0.5$, 1, 2, or 4 the relative error was always less than 1%. In fact, figure 5 panel (c) shows that most of the time the error was substantially less than 1%.

Interestingly, as remarked in subsection 1.3 QQMM itself is fairly computationally intensive and so that for "cheap" $V$'s, like the $L^{3/2}$ criterion employed in this section, one might expect that QQMM to be slower in terms of actual running time than, say, BFGS. However, it turns out that for all five choices of $\lambda$ except $\lambda := 0.1$ QQMM$_{rescue}$ actually had a shorter mean elapsed time than did BFGS$_{1,000,000}$. (Elapsed times were computed using the R function `system.time`, Becker *et al* (1, pp. 218–220).) This is particularly interesting in that `optim` must surely be a far more optimized program than is the still experimental QQMM.

### 3.3.1. The $\lambda := 2$ case

With $\lambda := 2$, the *maximum* number of evaluations required by QQMM$_{rescue}$ was 87. (In particular, QQMM never required "rescue" in this case.) The maximum required by BFGS$_{1,000,000}$ was 192. Thus, in this case the versions of the two optimizers were not comparable. So for $\lambda := 2$, I also compared QQMM to "$BFGS_{10}$" (BFGS restricted to a 10 iteration limit). $BFGS_{10}$ averaged slightly more evaluations (mean = 69) than did QQMM (mean = 53). (See Table 1.) But the table shows that *$BFGS_{10}$ was very inacccuarate,* especially compared to QQMM. The *smallest* relative error for $BFGS_{10}$ (compared to $BFGS_{1,000,000}$) was 12%! The *median* of the ratio

$$\frac{\text{relative error for } BFGS_{10}}{\text{relative error for QQMM}}$$

was *1787 (!).* (Here again "relative error" means error relative to $BFGS_{1,000,000}$.)

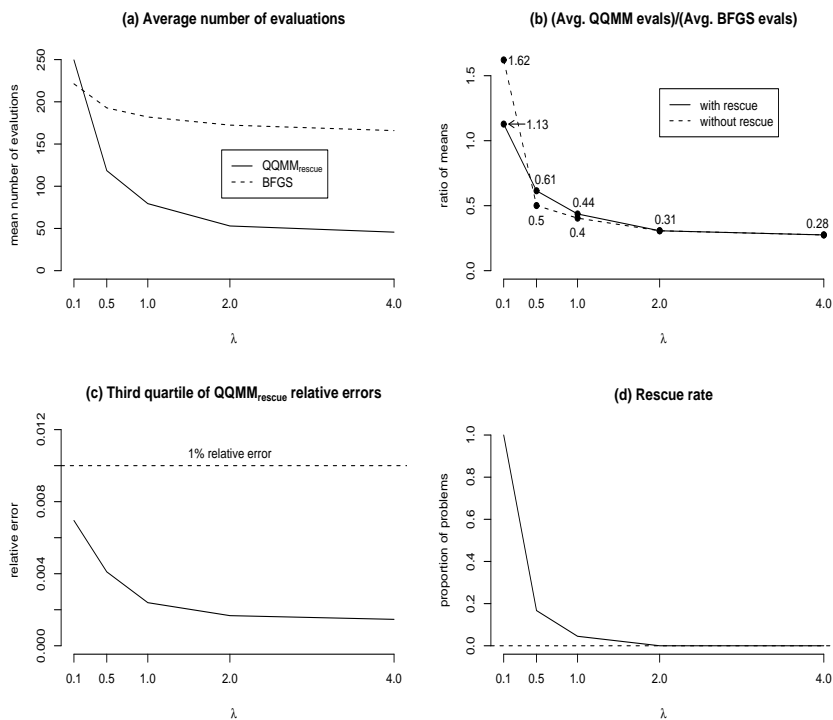

Fig 5. *Summaries of some simulation results. In each case the variable on the horizontal axis is the complexity parameter, $\lambda$. Panel (a) shows the mean absolute number of evaluations required by $QQMM_{rescue}$ (solid curve) and $BFGS_{1,000,000}$ (dashed curve; in the figure "BFGS" means "$BFGS_{1,000,000}$"). Panel (b) shows the ratio of the mean number of evaluations of $QQMM_{rescue}$ to that of $BFGS_{1,000,000}$ (solid curve) and QQMM (without rescue) to $BFGS_{1,000,000}$ (dashed curve). Panel (c) shows the third quartiles of the errors of $QQMM_{rescue}$ relative to $BFGS_{1,000,000}$. Panel (d) shows the proportion of simulations for which QQMM needed "rescue".*

| Optimizer | Avg. number of evaluations | Max. number of evaluations | Min. relative error | Median relative error |
|---|---|---|---|---|
| QQMM | 52.9 | 87.0 | $5 \times 10^{-4}$ | 0.0014 |
| $BFGS_{10}$ | 69.1 | 73.0 | 0.12 | 2.3 |

TABLE 1
*Some statistics from simulation with $\lambda := 2$. Note that here relative error is* not *expressed as a percent.*

## 4. Summary, Discussion, and Conclusions

Good, general optimizers for convex functions of the form (2), where $V$ is a known non-negative convex function and $Q$ is a known quadratic form, are apparently not readily available. QQMM is an apparently new optimizer designed for such problems. QQMM was designed to exploit the known structure of $F$ to find, with a small number of function evaluations, an approximate minimizer of $F$. QQMM makes use of two means, early stopping and careful choice of trial minimizers, to reach this goal.

A hybrid algorithm that employs "BFGS rescue" (subsection 2.5) was compared to the well regarded BFGS algorithm, the latter with a generous iteration limit. QQMM with BFGS rescue, denoted $QQMM_{rescue}$, means the following. *If* QQMM does not converge after a limited number of evaluations, BFGS is run for a limited number of iterations from where QQMM left off.

The comparison was based on fitting $L^{3/2}$ nonparametric regression models to simulated data. The regression algorithm employed a "complexity parameter", $\lambda$. To make the two optimizers comparable, I set the evaluation/iteration cutoffs for $QQMM_{rescue}$ so that, at least for small $\lambda$, the maximum number of evaluations of the hybrid was at least as large as that of BFGS. For production work different evaluation limits on QQMM and iteration limits on BFGS may be preferable. Moreover, with today's multicore computers another way to combine QQMM and BFGS presents itself: Run both QQMM, without rescue, and BFGS simultaneously! When one converges, stop the other one.

QQMM is an apparently new algorithm, so it was necessary to find some reasonable values for the tuning constants described in subsubsection 2.2.2 and subsection 2.3. This was done by systematically experimenting on tens of thousands of random two-dimensional optimization problems like the one illustrated in figure 1. (In particular, the tuning constants were chosen based on problems different from the regression problems that I used for testing in section 3.) Further work is needed in choosing good tuning constant values.

I argued that the mean number of evaluations needed for convergence was the proper outcome measure for assessing the performance of an optimizer in the context of interest. The reason is that QQMM is designed especially for problems in which function evaluation is expensive.

We found that, except for the lowest value of $\lambda$ tested, where $\text{QQMM}_{rescue}$ required 13% more evaluations, on average, than did BFGS, $\text{QQMM}_{rescue}$ required substantially fewer evaluations, on average, than did BFGS. Moreover, we found that in the example discussed in section 3, in which function evaluation was fast, usually $\text{QQMM}_{rescue}$ was faster than BFGS in terms of elapsed time as well as in number of function evaluations. I also tested QQMM without rescue, i.e., allowed to run until it converged. It was at least as fast as the rescued version except for the lowest value of $\lambda$, where it required substantially more evaluations than did BFGS.

Now, for the lowest value of $\lambda$ the algorithm over-fitted the data. In model fitting, one fits more models which seem to fit the data pretty well than one does fitting models that overfit. So based on our simulations one expects that in model fitting, a large majority of the time QQMM would out perform BFGS.

Might BFGS be improved to work well with the sort of problems studied here? It probably could. However, QQMM could then also be further refined and an "arms race" between BFGS and QQMM might ensue. The purpose of this paper is not to report on such an arms race. Instead, my goal here is merely to introduce the QQMM algorithm and give preliminary test results concerning its performance.

As remarked in subsubsection 2.2.2 and in subsubsection 3.1.3, QQMM requires the inverse of a matrix, $\mathbf{Q}$, defining $Q$. In the calculations described in this paper I improved the conditioning of $\mathbf{Q}$ by adding a small constant to each diagonal element of $\mathbf{Q}$. This procedure actually changes the optimization problem slightly. From a practical, statistical standpoint the solution to the approximate problem may be just as good as the solution of the exact problem.

However, it is worth investigating the possibility of a version of QQMM that does not require $\mathbf{Q}^{-1}$. In the meantime, one easy fix for this problem is to *always* follow up QQMM with a short BFGS run, even when QQMM does not require "rescue", starting from where QQMM left off but applying BFGS to the original problem with the unmodified $\mathbf{Q}$. A similar strategy could also be applied when $V$ is unbounded below and so, as suggested in subsection 1.2, is replaced by $\max\{V - c, 0\}$ for some constant $c$.

Another improvement to $\text{QQMM}_{rescue}$ might be possible. For the BFGS rescue we started BFGS from where QQMM left off, presumably in some rough neighborhood of the true minimizer. However, one can imagine using the QQMM run to provide even more information to BFGS. BFGS uses a matrix, call it $\mathbf{B}$, that functions like the inverse Hessian in Newton's method. BFGS computes that matrix recursively as it evaluates $F$ and its gradient. One could imagine having QQMM give BFGS a further head start by having QQMM compute an initial **B** for BFGS based on *its* (i.e., QQMM's) evaluation history. So far, my experiments with this procedure have not proved successful. However, I believe this idea is worth further work.

I began this paper (subsection 1.1) by mentioning my efforts to develop a nonparametric hazard function estimator. My impression was that QQMM was much faster than BFGS for hazard estimation. This nonparametric estimator is, at this writing, unpublished. However, if I may, I conclude this paper by describing what happened the one and only time I used both BFGS and QQMM for the same hazard estimation problem. (This obviously has only anecdotal bearing on the question of the relative speeds of the two methods.) The comparison was made in estimating the hazard from a real data set. The kernel and complexity parameter for $F$ as in (1) were such that the model fit well. (In particular, the estimator was tuned so that it did not overfit.) The number of variables in the optimization was $d = 2508$. QQMM (without rescue but tuned as in section 3) took less than 30 minutes. BFGS took almost 5.3 hours. So BFGS took over 11 times longer than QQMM.

However, one might wonder how long BFGS would take had it been stopped more efficiently. Define "efficient stopping" to mean stopping as soon as a value of the objective function had been reached that was less than 1% more than the minimum value of the objective function. ($\epsilon := 1\%$ in the QQMM run.) I ran BFGS again using the best possible stopping rule, *viz.*, one based on an "oracle". (An "oracle" is a source of information not ordinarily available in practice.) The 5 hour BFGS run provided a good estimate of the minimum value of the objective. From that the value one can accurately compute the value 1% more than the minimum. Again, the 5 hour BFGS run revealed that BFGS first reached below the 101% threshold on the $180^{th}$ iteration. So I ran BFGS again this time with an iteration limit of 180. With this stopping rule BFGS ran for a little over 2 hours before stopping. Thus, in this particular "real world" instance, QQMM *without* an oracle ran almost 4.5 times faster than BFGS did *with* an oracle. (QQMM computes an upper bound on the relative error, but without an oracle it cannot compute the true relative error.)

This anecdote and the experiment with early stopping of BFGS in the $L^{3/2}$ estimation problem with $\lambda := 2$ (subsubsection 3.3.1) suggest that the reason that QQMM is ordinarily faster than BFGS does not merely lie in the fact that QQMM allows good control of stopping (so stopping early is possible). QQMM generally makes the objective function decrease faster than does BFGS. (Of course, QQMM is specially designed for problems of the form (2) with convex $V$ bounded below by a known value and known $Q$. BFGS can handle a much broader range of problems. )

Based on the numerical experiments described in section 3, experiments with tens of thousands of random simulation problems like that shown in figure 1, and applications of QQMM to survival analysis mentioned in subsection 1.1 and above, QQMM seems to meet its goals rather well.

**Acknowledgments**

This paper benefits from discussion with Michael Overton and Michael C. Grant.